\documentclass[12pt]{article}
\newcommand{\bsigma}{\mbox{\boldmath $\sigma $}}

\begin{document}

%
%
%
%
%
%
%
%
%
\begin{titlepage}
\title
 {From Clifford Algebra of Nonrelativistic Phase Space to Quarks and Leptons of the Standard Model \footnote{Presented at the 10-th International Conference on Clifford Algebras and their Applications in Mathematical Physics (Tartu, Estonia, 2014); to appear in {\it Advances in Applied Clifford Algebras}.}}

\author{
{Piotr \.Zenczykowski \footnote{email: piotr.zenczykowski@ifj.edu.pl}}\\
{\it Division of Theoretical Physics}\\
{\it The Henryk Niewodnicza\'nski
Institute of Nuclear Physics}\\
{\it Polish Academy of Sciences}\\
{\it PL 31-342 Krak\'ow, Poland}
}


\maketitle
\begin{abstract}
We review a recently proposed Clifford-algebra approach to elementary particles. We start with: (1) a philosophical background that motivates a maximally symmetric treatment of position and momentum variables, and: (2) an analysis of the minimal conceptual assumptions needed in quark mass extraction procedures. With these points in mind,
a variation on Born's reciprocity argument provides us with an unorthodox view on the problem of mass. The idea of space quantization suggests then the linearization of the nonrelativistic quadratic form 
${\bf p}^2 +{\bf x}^2$ 
with position and momentum satisfying standard commutation relations. This leads to the 64-dimensional Clifford algebra
${Cl}_{6,0}$ 
of nonrelativistic phase space within which one identifies the internal quantum numbers of a single Standard Model generation of elementary particles (i.e. weak isospin, hypercharge, and color). The relevant quantum numbers are naturally linked to the symmetries of macroscopic  phase space. It is shown that the obtained phase-space-based description of elementary particles gives a subquark-less explanation of the celebrated Harari-Shupe rishon model. Finally, the concept of additivity is used to form novel suggestions as to how hadrons are constructed out of quarks and how macroscopically motivated invariances may be restored at the hadron level.
\end{abstract}

\end{titlepage}
\section{Introduction}
The ideas presented below were developed over the last several years in a series of texts 
\cite{ZenAPPB1,ZenJPA,ZenBook}. For the lack of space, here we limit ourselves to the  
most important points of our approach only.  First, we will stress the conceptual basis of our scheme, in particular  the philosophical need for a maximally symmetric treatment of position and momentum coordinates.  This fundamental argument and the analogy with
 the Dirac linearization procedure will lead us to the Clifford algebra ${Cl}_{6,0}$ as 
a structure that should naturally describe some aspects of nonrelativistic phase space. Second, a discussion of a possible, rotationally noninvariant generalization of the concept of mass will be given. The discussion will involve both phenomenology and position-momentum symmetry arguments. 
Third, we will show that the Clifford algebra in question forms
a vehicle that satisfactorily describes the most salient features of a single generation of elementary particles in the Standard Model (SM). It will also be shown that this algebra provides a  phase-space-related explanation of the Harari-Shupe (HS) rishon model of elementary particles, an explanation that avoids all the shortcomings of the HS model itself. 
Fourth, ways of restoring rotational (and translational) invariances at the level of individually observable particles composed of quarks (i.e. hadrons) will be briefly discussed. For details, the interested reader is kindly referred to the original papers.

\section{The Philosophical Argument}
The Standard Model of elementary particles is extremely successful. It organizes experimental results on the structure of matter into a single framework that provides us with a very useful idealization of certain aspects of physical reality. The SM field-theoretical idealization is based on the Democritean philosophy which imagines reality as built of indivisible `atoms' (elementary particles) moving in continuous background space.

Yet, the Standard Model does not answer many simple and important questions that are usually asked by the non-experts. For example, it does not state why elementary particle masses have the particular values they have, or what is the origin of the $U(1)\otimes SU(2) \otimes SU(3)$ gauge group built into the model to describe the three types of fundamental interactions (electromagnetic, weak, and strong). In fact, the values of particle masses or the type of gauge group are put into the SM solely on the basis of experimental findings. We obviously lack a deeper conceptual principle.

Since the Standard Model, as any theory of ours, provides an idealized (and therefore limited) description of certain aspects of physical reality only, finding an answer to such questions may in fact require dispensing with the field-theoretical idealization itself and its Democritean philosophical background.
Thus, in order to proceed, one might need an extension of the Democritean view of nature as being built of `things' moving in background space as time `flows'. 
In fact, various philosophers have argued that one has to consider not only `things' but also `processes', and that they should be introduced into philosophical systems as two parallel and independent concepts.  Such ideas were expressed in various ways by Heraclitus, Mach, Whitehead,  Barbour \cite{Barbour,Whitehead} and many other philosophers and physicists. For example, Mach argued in favor of replacing the abstract concept of time with the underlying concept of correlated `change', while Whitehead stressed that 'the flux of things is one ultimate generalization around which we must weave our philosophical system'.

It is also worth stressing that while the Whiteheadian process 
philosophy preceded the appearance of the full-fledged quantum theory,
several of its philosophical concepts may be put into a strict correspondence with physical quantum concepts that were formed only later. For example, the Whiteheadian concept of `actual occasion' corresponds naturally to the Heisenbergian concept of `actualization'
(or the collapse of the quantum state).
Consequently, Whiteheadian philosophical arguments could have been  regarded as a philosophical anticipation of certain aspects of quantum physics. 
One may therefore claim that deep philosophical arguments suggest a generalization of the Democritean standpoint to an approach in which things and processes are treated in the most symmetric way possible.
In other words, the notion of process should be viewed not in the simplistic Democritean way (i.e. not as a derived concept) but in a more Heraclitean manner (i.e. as a fundamental notion).
The translation from the philosophical language of things and processes to the physical one is simple: things correspond to their (static) positions, while processes --- to their momenta (which define changes in these positions).

\section{The Problem of Mass}
In the middle of 20th century Max Born expressed similar longing for the most symmetric treatment of positions and momenta \cite{Born}.
He observed that various physical formulas, such as the Hamilton's equations of motion or the position-momentum commutation relations, exhibit `reciprocity symmetry', i.e. an exact symmetry under the interchange:
\begin{eqnarray}
{\bf x} \to {\bf p}, & ~~~~~~~ & {\bf p} \to {\bf - x}. 
\end{eqnarray}  
Born observed that the concept of mass breaks this symmetry totally (i.e. that quantized mass is always associated with momentum, never with position coordinates). He was deeply dissatisfied with this state of affairs and concluded: `This lack of symmetry seems to me very strange and rather improbable'. 

Born's reciprocity idea was one of many attempts to attack the problem of mass and to introduce some generalization of this concept. Today, the elementary particles of the Standard Model are endowed with mass using the Higgs mechanism. Yet, putting aside all the propaganda involved, this mechanism does not bring us much closer to the resolution of the problem of mass: essentially, it only shifts the problem. The pattern of lepton and quark masses remains completely unexplained. The observed quantization of elementary particle masses seems to indicate that the problem should be treated in parallel to the quantization of spin, i.e. that in its essence it is an algebraic problem.
Unfortunately, so far nobody has come with a working idea how to do that.

On the other hand, over half a century after Born's lament we have amassed a lot of experimental information that was unknown to him. In particular, we have identified two types of fundamental matter particles: leptons and quarks. The mass of any lepton is easily measurable in experiments as leptons are individually observable particles which move freely over macroscopic distances.  For quarks, which have not been observed in asymptotic states, the concept of mass is much more nebulous.

The standard picture presents quarks as tiny particles `moving freely' at small distances within hadrons according to the field-theoretical rules, but undergoing strong confining interactions when interquark distances become large. Clearly, this picture assumes that the Democritean concept of continuous background space may be rightfully extrapolated to the interior of hadrons. Yet, such an extension need not be correct: the concept of spacetime is operationally defined via the use of {\it macroscopic} rods and clocks. There is no way to fit such rods or clocks into any hadron \cite{Wigner}. Thus, the assumption of an ordinary background space within hadrons may be justified only a posteriori, i.e. only after the theoretical calculations of hadronic properties exhibit fine agreement with the results of hadronic experiments. To summarize, in the case of quarks we should look for experimental clues that would clarify the issue of the purported existence of ordinary background space within hadrons.

The first related point we want to make here is connected with the question: how one can define (and estimate) the mass of an object called quark if free quarks are never observed. Obviously, quark mass must be extracted from some experimental data using some theoretical assumptions. The question is `what are the minimal theoretical assumptions needed for such an extraction to be acceptable?' The answer to that question is provided by the Gell-Mann--Oakes-Renner (GMOR) formula \cite{GMOR} which reads:
\begin{equation}
m^2_{\pi}\delta^{ab}=-\frac{1}{f^2_{\pi}}\langle 0| [Q^a_5,[Q^b_5,H(0)]]|0\rangle.
\label{GMOR}
\end{equation}
Here, $m_{\pi}$ and $f_{\pi}$ are the (directly determined) pion mass and decay constant, $a,b$ are isospin indices, $Q^a_5$ is an axial charge, $H(0)$ is the Hamiltonian taken at $x=0$, and $|0\rangle $ represents hadronic vacuum. If one inserts the standard quark mass term $m_q \bar{q}q$ into $H(0)$ one can readily evaluate the double commutator to obtain (using the fact that pion is composed of quarks $q= u,d$):
\begin{equation}
m^2_{\pi}=-\frac{1}{f^2_{\pi}}(m_u\langle 0| \bar{u}u|0\rangle +
m_d \langle 0|\bar{d}d|0\rangle).
\end{equation}
Assuming $u\leftrightarrow d$ symmetry for the vacuum expectation values $\langle 0 | ... |0 \rangle$ and extending it to the strange $(s)$ quark one can divide out these expectation values and obtain the ratios of the so-called `current' quark masses given in terms of the ratios of pion and kaon masses squared. This is how the often mentioned two ratios  of quark masses, i.e.
\begin{eqnarray}
\frac{2m_s}{m_u+m_d} \approx 25, & ~~~{\rm and}~~~ & \frac{m_u}{m_d} \approx 0.56,
\end{eqnarray}
are extracted from experiment. The absolute values of $m_u,m_d,m_s$ are then determined by observing that changing a nonstrange ($u,d$) quark into a strange one adds about 150 MeV to the mass of a baryon (hence $m_s \approx m_u+150~MeV \approx m_d+150~MeV$). In this way rough values of quark masses are extracted from the data. The lattice QCD evaluation of quark masses \cite{PDG}, which basically agrees with this approach, differs in that it adds many more details to it. 

The crucial point underlying our ideas on the concept of mass is that the GMOR extraction of quark masses {\it does not use} the notion of quark momentum at all,  using only concepts defined at the hadronic level (see Eq. (\ref{GMOR})). In that it is very different from the QCD-based approach which involves the standard picture of quarks moving `within' hadrons according to the field-theoretical rules. Thus, the GMOR-extracted values of quark masses are more general than those obtained in QCD: they admit not only QCD, but also a wider spectrum of quark confining theories. This point is very important since there are hints from baryon spectroscopy that something is missing in the original quark model/QCD approach. The problem that we allude to is called the problem of {\it missing excited baryons}.
Namely, the standard quark model (as well as lattice QCD) predicts the existence of many more excited baryonic states than experimentally observed. Specifically, at the $N=3$ level of the standard harmonic oscillator approach to the baryon spectrum there are many predicted $SU(6)\otimes O(3)$ multiplets: $(56,0^+)$, $(70,0^+)$, $(56, 2^+)$, $(70, 2^+)$, $(20, 1^+)$. The list of experimentally observed $N=3$ states is much shorter. In particular, no $(20, 1^+)$ multiplet is seen.  There are two ways of resolving the problem: either there is a frozen spatial degree of freedom in baryons, or the missing baryons decay into channels that are not easily observable experimentally.
The situation is so serious that Capstick and Roberts \cite{CapstickRoberts} write: {\it `These questions about baryon physics are fundamental. If no new baryons are found, both QCD and the quark model will have made incorrect predictions, and it would be necessary to correct the misconceptions that led to these predictions. Current understanding of QCD would have to be modified and the dynamics within the quark model would have to be changed.'} 

The above discussion sets the stage for the phenomenological acceptability of a generalized concept of mass, to be presented shortly.
Yet, before we do just that, we have to defend our approach against a simple charge. Namely, in the previous section we argued for a maximally symmetric treatment of nonrelativistic position and momentum coordinates. This  begs a question: where is special relativity? We argue that at the beginning stages of the construction of a phase-space-related approach it is fully justified to restrict the considerations to a nonrelativistic approach. 
First, known classical-quantum connections between the properties of macroscopic space and the spatial quantum numbers may be established by applying strictly nonrelativistic considerations.
Indeed, spin is related to 3D rotations, parity --- to 3D reflection, and the existence of particles and antiparticles --- to the reversal of nonrelativistic time.\footnote{In fact, the linearization of nonrelativistic Schr\"odinger equation does lead to antiparticles.}
Second, the quantum-classical tension exists not only between general relativity and quantum physics (as exemplified by problems with the quantization of gravity), or between special relativity and quantum physics (as Bell's theorem and the nonlocality of quantum theory suggest), but also between nonrelativistic classical physics and nonrelativistic quantum theory. Indeed, the classical physics and the quantum physics differ dramatically already at the nonrelativistic level. We conclude that the idea of extending the standard concept of mass and connecting it with other (internal) quantum numbers should be first attempted at the nonrelativistic level. We come back to the issue of special relativity at the end of the paper.

Let us now return to the Born's conception. We recall that Max Born wanted to interchange the momenta and positions via a reciprocity transformation (${\bf x} \to {\bf p}$, ${\bf p} \to - {\bf x}$). In general, however, not two but eight different choices can be made for the canonical positions and momenta, i.e.:
\begin{eqnarray}
({\rm can.~position}, ~ {\rm can.~momentum})&~~~~~~~~&({\rm can.~position}, ~ {\rm can.~momentum})\nonumber\\
(x_1,x_2,x_3), ~~~~~ (p_1,p_2,p_3)~~~~~~~~&&~~~~~(p_1,p_2,p_3), ~~~~~ (x_1,x_2,x_3)\nonumber \\
(x_1,p_2,p_3), ~~~~~ (p_1,x_2,x_3)~~~~~~~~&&~~~~~(p_1,x_2,x_3), ~~~~~ (x_1,p_2,p_3)\nonumber\\
(p_1,x_2,p_3), ~~~~~ (x_1,p_2,x_3)~~~~~~~~&&~~~~~(x_1,p_2,x_3), ~~~~~ (p_1,x_2,p_3)\nonumber\\
(p_1,p_2,x_3), ~~~~~ (x_1,x_2,p_3)~~~~~~~~&&~~~~~(x_1,x_2,p_3), ~~~~~ (p_1,p_2,x_3).
\label{weird}
\end{eqnarray}
The two sides of the first line above are connected (up to a sign) via the reciprocity transformation of Max Born, as are the left- and right-hand sides of the remaining three lines
(the reciprocity transformation does not link the four lines of Eq. (\ref{weird})). The latter three lines represent new possibilities with which three variations on the concept of mass could be associated.
The four possibilities on the l.h.s. of Eq. (\ref{weird}) correspond to even permutations of phase-space variables (and are not equivalent among themselves), while those on the r.h.s. --- to odd permutations. Violation of the reciprocity transformation by the concept of mass suggests that this concept should also violate the general transitions from the even to the odd permutations. Thus, one should not expect the concept of mass to be related to canonical momentum $(x_1,p_2,p_3)$ (etc.). However, what about the three even permutations on the l.h.s. above? Obviously, they violate rotational (and translational) invariances. Therefore, the related masses should exhibit similar features. Yet, these properties do not have to be deadly for a physical theory. The only condition we really have to meet is that the rotational (and translational) invariances must be recovered in our macroworld. It may happen that the weird objects associated with lines two, three and four on the l.h.s. of Eq. (\ref{weird}) conspire among themselves and form conglomerates that behave properly from the rotational and translational point of view. Accordingly, we put forward a far reaching conjecture that the 1+3 possibilities on the l.h.s. above correspond to the existence of leptons and three-colored quarks. 

We will show shortly that in fact these 1+3 possibilities may be naturally connected with the standard internal quantum numbers characterizing leptons and quarks of the Standard Model. First, however, we have to stress that the whole discussion above refers to the concept of mass and internal quantum numbers only. Thus, the field-theoretical description of quarks by quark fields $q(x)$ is supposed to be completely unaffected by our considerations {\it as long as the concept of mass is not introduced}.
The deviations should appear only in those places where standard quark propagators (involving the standard concept of mass) should appear. Indeed,  for a quark, the momentum of a lepton should get replaced by one of the weird canonical momenta on the l.h.s. of Eq. (\ref{weird}).

\section{Clifford Algebra of Nonrelativistic Phase Space}
We start with the observation that the quantum concept of spin may be
arrived at by the linearization of the 3D invariant ${\bf p}^2$ (via ${\bf p}^2=({\bf p}\cdot{\bsigma})({\bf p}\cdot{\bsigma})$). 
The most ${\bf x} \leftrightarrow {\bf p}$ symmetric extension  of this argument suggests then the linearization of the expression
${\bf p}^2+{\bf x}^2$. 
(We will hint at the possible relativistic extension of the proposed scheme at the end of the paper.)
Since position and momentum do not generally commute, this linearization has to be performed subject to the condition that
$[x_j,p_k]=i\delta_{jk}$. Accordingly, one finds that the square of  expression ${\bf A}\cdot{\bf p}+{\bf B}\cdot{\bf x}$, with ${\bf A}$ and ${\bf B}$ being six mutually anticommuting objects, obeys the formula:
\begin{equation}
({\bf A}\cdot{\bf p}+{\bf B}\cdot{\bf x})({\bf A}\cdot{\bf p}+{\bf B}\cdot{\bf x})={\bf p}^2+{\bf x}^2 +R,
\label{square}
\end{equation}
where an additional term, i.e. $R$, appears because position and its conjugated momentum do not commute. The anticommuting objects
${\bf A}$ and ${\bf B}$ may be represented by eight-dimensional matrices:
\begin{eqnarray}
A_k&=&\sigma_k\otimes\sigma_0\otimes\sigma_1,\nonumber \\
B_j&=&\sigma_0\otimes\sigma_j\otimes\sigma_2.
\end{eqnarray}
One easily finds that
\begin{eqnarray}
R=-\frac{i}{2}\sum_k[A_k,B_k]&=&\sum_k\sigma_k\otimes\sigma_k\otimes\sigma_3.
\end{eqnarray}
The seventh anticommuting element
of the Clifford algebra in question is denoted as
\begin{eqnarray}
B=iA_1A_2A_3B_1B_2B_3&=&\sigma_0\otimes\sigma_0\otimes\sigma_3.
\end{eqnarray}

We define now 
\begin{eqnarray}
I_3=\frac{1}{2}B,&~~~~~~~~ &Y=\frac{1}{3}RB,
\end{eqnarray}
and observe that $I_3$ and $Y$ commute with the operators describing ordinary 3D rotations and 3D reflections in phase space.  The eigenvalues of $I_3$ and $Y$ are:
\begin{eqnarray}
I_3 \to \pm \frac{1}{2}, & ~~~~~~ & Y \to 
-1,+\frac{1}{3},+\frac{1}{3},+\frac{1}{3}.
\end{eqnarray}
Since 
 $I_3$ and $Y$ are obviously different 
from spin, parity, or charge conjugation parity, they constitute candidates for two new (phase-space-related) quantum numbers. The basic conjecture is that
Eq. (\ref{square}) (when appropriately modified) should be
identified with the Gell-Mann -- Nishijima formula for charge $Q$:
\begin{equation}
Q \equiv \frac{1}{6}\left[ ({\bf p}^2 +{\bf x}^2)_{vac}+R \right] B =
I_3+\frac{Y}{2},
\end{equation}
where the subscript $vac$ denotes the lowest (`vacuum') eigenvalue of ${\bf p}^2+{\bf x}^2$, i.e. $3$, while $I_3$ and $Y$ are weak isospin and hypercharge, respectively. Indeed, the eigenvalues of $Q$ are: 
$(0,+2/3,+2/3,+2/3,-1,-1/3,-1/3,-1/3)$, i.e. they are identical with
the charges of 8 fundamental particles from a single generation of the Standard Model.

\subsection{The structure of hypercharge and the rishon model}
The hypercharge $Y$ of lepton and three colored quarks is built out of three mutually commuting `partial hypercharges" $Y_k = -\frac{i}{6}[A_k,B_k]$  ($k=1,2,3$) in the way shown in Table \ref{tbl1}.
\begin{table}[h]
\caption{Hypercharge Y and its component partial hypercharges $Y_k$}
\begin{center}
{\begin{tabular}{cccc} \hline
 $~Y_1 $&$~Y_2$&$~Y_3$&$~~Y$
\rule{0mm}{6mm}\\
\hline
$-\frac{1}{3}$&$+\frac{1}{3}$&$+\frac{1}{3}$&$+\frac{1}{3}$\rule{0mm}{6mm}\\
$+\frac{1}{3}$&$-\frac{1}{3}$&$+\frac{1}{3}$&$+\frac{1}{3}$\rule{0mm}{6mm}\\
$+\frac{1}{3}$&$+\frac{1}{3}$&$-\frac{1}{3}$&$+\frac{1}{3}$\rule{0mm}{6mm}\\
$-\frac{1}{3}$&$-\frac{1}{3}$&$-\frac{1}{3}$&$-1$\rule{0mm}{6mm}\\
\hline
\end{tabular}}
\end{center}
\label{tbl1}
\end{table}

The pattern shown in Table \ref{tbl1} is in one-to-one correspondence   with the way in which elementary particle charges are  built out of `subquark' charges in the celebrated Harari-Shupe (HS) rishon model
\cite{HarariShupe}. In that model, quarks and leptons are composed of only two `truly fundamental' spin-1/2 particles, the `rishons': $T$ of charge $+1/3$, and $V$ of charge $0$ (together with their antiparticles $\bar{T}$ and $\bar{V}$), as shown in Table \ref{HSM}.
The correspondence between the HS model and the phase-space approach is
\begin{eqnarray}
Y_k=-1/3 \leftrightarrow V, &~~~~~~& Y_k=+1/3 \leftrightarrow T,
\end{eqnarray}
where $k$ labels the position in the ordering of rishons in Table \ref{HSM}. Although the HS model is very economic, it exhibits many shortcomings. Indeed, it predicts the existence of various unobserved particles such as  spin-3/2 partners of leptons and quarks or the $TT\bar{T}$ states. Furthermore, the expected SU(3) color symmetry does not really appear in the model, the rishons are not antisymmetrized in spite of being fermions, the model predicts unobserved baryon-number violating processes, etc.
\begin{table}[h]
\caption{The Harari-Shupe Model: rishon structure of the $I_3=+1/2$ members of a single SM generation}
\begin{center}
{\begin{tabular}{cccccccc} \hline
 $\nu_e $&$u_R$&$u_G$&$u_B$
&$e^+ $&$\bar{d}_R$&$\bar{d}_G$&$\bar{d}_B$\rule{0mm}{6mm}\\
 $\displaystyle VVV$&$\displaystyle TTV$&
$\displaystyle TVT $&$\displaystyle VTT$&$TTT$&$VVT$&$VTV$&$TVV\vphantom{\frac{1}{j_k}}$\rule{0mm}{6mm}\\
\hline
\end{tabular}}
\end{center}
\label{HSM}
\end{table}

Therefore,  it has to be stressed here that, despite the correspondence with the HS model, the phase-space Clifford-algebra approach achieves much more than just reproducing this model. Namely, it turns out that the shortcomings of the HS model simply do not appear in the phase-space approach. This is basically because the Clifford algebra approach does not introduce subparticles, but deals with phase-space symmetries only (the components of charge operator are not associated with subparticles). For details, see
\cite{ZenAPPB1,ZenJPA,ZenBook}.

\subsection{Even subalgebra of Clifford alegbra}
The even subalgebra of the phase-space Clifford algebra consists of $16+ 16$ elements, the first and second set working in sectors $I_3=+ 1/2$ and $-1/2$, respectively. Each one of the 16-element sets decomposes into a unit element and 15 generators of $SO(6)$ in a given sector. The set of generators decomposes in $SU(3)$ as $15=1\oplus 8\oplus 3 \oplus 3^*$. The $3$ and $3^*$ generate lepton-quark transformations. For example, a rotation by $\pi/2$ generated by $F_{+2} \equiv -\frac{i}{4} \epsilon_{2kl}[A_k,B_l]$ swaps quark \# 2 with a lepton while leaving quarks \# 1, 3 untouched \cite{ZenAPPB1}. The corresponding transformation in phase space is
\begin{equation}
(p_1,p_2,p_3)\to(-x_1,p_2,x_3).
\end{equation}
Thus, Clifford algebra provides a strict connection between the phase-space-based canonical momentum (and mass) heuristic as anticipated in Eq. (\ref{weird}) and the structure of internal quantum numbers as presented in Table \ref{tbl1}.
This is a highly non-trivial connection. From the point of view herein advocated, the three colored quarks should be regarded as leptons whose canonical momenta (and mass) are `rotated' in phase space in three possible ways. 

\subsection{Odd part of Clifford algebra}
The odd part of Clifford algebra consists of 16+16 elements which link sectors of $I_3 =\pm 1/2$. Such a 16-element set decomposes in $SU(3)$ as $16 = 1\oplus 3 \oplus 3 \oplus 3^* \oplus 6$. One can check that the $SU(3)$ singlet  (denoted $G_0$) works in the lepton subspace $Y = -1$ only.
Its explicit form is
\begin{equation}
G_0 \propto (1-\sum_k\sigma_k\otimes\sigma_k)\otimes (\sigma_1+i\sigma_2).
\end{equation} Since the only $SO(3)$-scalar odd elements of Clifford algebra are
 $G_0$ and its hermitian conjugate, it is only from them that a candidate for an algebraic counterpart of the mass term of a lepton may be formed.

One can similarly check that the $SU(3)$ sextet (denoted $G_{\{kl\}}$) works in the quark subspace $Y=+1/3$ only. In fact, the previously discussed finite rotation generated by $F_{+2}$ (that swaps lepton and quark \# 2) transforms $G_0$ into 
\begin{equation}
G_{22}\propto (1+\sum_k\sigma_k\otimes\sigma_k-2\sigma_2\otimes\sigma_2)\otimes(\sigma_1+i\sigma_2).
\end{equation} 
Thus, $G_{22}$ and its hermitian conjugate constitute elements from which an algebraic counterpart of the mass term of quark \# 2 could be constructed.
Since only the trace of $G_{\{kl\}}$ is a rotational scalar, the $G_{22}$-induced mass term is not rotationally invariant, in agreement with our earlier heuristic arguments. Rotational invariance is, however, restored for the trace $\sum_kG_{kk}$, as originaly expected. This is an example of the expected quark conspiracy mechanism.
There are two corrections that have to be made for the above conspiracy conjecture to work. First, one obviously has to work with algebraic mass counterparts that are hermitian combinations of $G_0$ and $G_0^{\dagger}$ (or $G_{22}$ and $G_{22}^{\dagger}$). This does not present any difficulties.
The second, more difficult point appears when one attempts to restore special relativity. We will comment on it later. 

\subsection{Additivity and its consequences}
An important question for the phase-space approach to quarks is whether it can be used for the description of colorless composite states. Unfortunately, we have no detailed idea how to construct such a scheme. Still, some expectations may be presented. These expectations are based on the concept of additivity.

\subsubsection*{Additivity of canonical momenta}
 Specifically, we observe that a trivial but important concept in all of the standard physics is that of the additivity of the momenta of  component particles. A system of leptons, or a system of hadrons ($A,B,...$) is assigned a total momentum which is built as a sum of all the momenta of its components: ${\bf p}_{tot}={\bf p}_A + {\bf p}_B+...$~. Yet, for the quarks of the phase-space scheme, some of the components of ordinary momenta have been replaced with the components of positions, thus forming the canonical momenta of these quarks. It seems then natural to extend the additivity of physical momenta of ordinary particles to the additivity of canonical momenta of quarks.
An analysis of this prescription shows \cite{ZenJPA,ZenBook} that one can form translationally invariant expressions for systems composed of a quark-antiquark pair ($q\bar{q}$), or of three quarks of different colors ($qqq$), but not for two quarks $qq$, nor for $qq\bar{q}$ states, etc. Thus a result very similar to that obtained in the standard group-theoretical $SU(3)_{color}$-singlet explanation of quark confinement is predicted. In addition, the phase-space picture provides hints that for baryons this type of additivity leads to a frozen degree of freedom \cite{ZenBook}.

\subsubsection*{Additivity of quark Hamiltonians}
Another place where the concept of additivity seems to play an important role is in the construction of a rotationally and translationally invariant effective quark Hamiltonian. Let us restrict ourselves to the discussion of the canonical momentum part of such Hamiltonians. 
For a lepton we have the standard form
\begin{equation}
H={\bf A}\cdot{\bf p}.
\end{equation}
Transition to a quark of given color is achieved via a corresponding rotation in phase space. For quark of color \# 1, this leads to
$H^{(1)}=A_1p_1+B_2x_2+B_3x_3$. We observe that a change 
$(A_3,B_3) \to (-A_3,-B_3)$ does not affect $I_3$ or $Y$.
Thus, for the canonical momentum part the quark Hamiltonian 
one may take:
\begin{equation}
H^{(1)}=A_1p_1+B_2x_2-B_3y_3,
\label{quark1}
\end{equation}
where we have renamed $x_3$ as $y_3$ so that in a formula that follows the position coordinates of different quarks are named differently. 
The Hamiltonians of quarks of colors \# 2 and \# 3 are obtained by cyclic permutation.
Summation of the three Hamiltonians yields the formula for an effective quark Hamiltonian:
\begin{equation}
H^{eff}=\sum_k H^{(k)} = A_1p_1 + A_2p_2 + A_3p_3 +
B_1(x_1-y_1) + B_2(x_2-y_2) + B_3(x_3-y_3).
\end{equation}
Thus, summation over quark colors leads to a
 translationally invariant effective expression. If one further assumes that quarks of different colors are located at the same point in space (i.e. $x_k=y_k$), the above expression simplifies to:
\begin{equation}
H^{eff}= {\bf A} \cdot {\bf p},
\label{rotinv}
\end{equation} 
which looks exactly like the momentum part of the standard lepton Hamiltonian. The difference when compared to the standard approaches is that for quarks of the phase-space approach such a lepton-like form is obtained not for a quark of a fixed color, but for an expression that involves {\it a sum over colors}. The appearance of form (\ref{rotinv}) at the level of the summed-over-color expressions constitutes a minimum requirement in the construction of a physically viable theory. Indeed, the rotational and translational invariances certainly have to be recovered for colorless objects coupling to the color-blind probes  (photons, weak bosons) that are used to study quarks.

\subsubsection*{Mass and special relativity}
So far, our discussion was nonrelativistic. Obviously, one has to bring special relativity in. As with the case of rotational invariance, it has to appear at the level of color-blind objects, i.e. at the level of $H^{eff}$ (the ordinary concept of relativistic spacetime has to be recovered at the level of colorless hadrons).
More specifically, the momentum part of the effective Hamiltonian 
(\ref{rotinv}) has to be supplemented with a quark mass part, so that
(after squaring the relevant linearized expression)
a relativistically motivated expression ${\bf p}^2+m^2$ might appear at an effective color-blind quark level. 
When trying to do that within the 64-element Clifford algebra discussed so far, one encounters an obstacle: the candidate effective quark mass term, built from $\sum_kG_{kk}$ and  $\sum_kG^{\dagger}_{kk}$, works in the same subspace as the algebraic momentum counterparts $A_m$. This conflicts with the construction of the relativistically motivated expression ${\bf p}^2+m^2$. 
Preliminary studies \cite{ZenInPrep} indicate that the problem may be solved by an appropriate extension of the Clifford algebra in question, in a way similar to the original Dirac's treatment of the momentum and mass terms. As with rotational covariance before, special relativity becomes then recovered only after the summation over quark colors is performed. This differs from the standard approach in which 
the requirements of rotational and relativistic covariances are assigned to individual colored quarks.


\end{document}